\documentclass[sigconf,unicode,bookmarks=false]{acmart}
\usepackage{xcolor}
\usepackage[inline]{enumitem}
\usepackage{verbatim}
\usepackage{booktabs} 
\usepackage[capitalise,noabbrev]{cleveref} 
\usepackage{tabularx}
\usepackage{xcolor,colortbl}
\usepackage{comment}
\usepackage{enumitem}
\usepackage{soul}

\newcommand{\multicell}[2][t]{\begin{tabular}[#1]{@{}l@{}}#2\end{tabular}} 
\newcommand{\boxbox}[2]{
    \begin{center}
    \fbox{
        \parbox{#1\columnwidth}{#2
        }
    }
    \end{center}
}
\newcommand{\ra}[1]{\renewcommand{\arraystretch}{#1}}

\AtBeginDocument{%
  \providecommand\BibTeX{{%
    \normalfont B\kern-0.5em{\scshape i\kern-0.25em b}\kern-0.8em\TeX}}}

\copyrightyear{2020}
\acmYear{2020}
\setcopyright{acmlicensed}
\acmConference[ICSEW'20]{IEEE/ACM 42nd International Conference on Software Engineering Workshops}{May 23--29, 2020}{Seoul, Republic of Korea}
\acmBooktitle{IEEE/ACM 42nd International Conference on Software Engineering Workshops (ICSEW'20), May 23--29, 2020, Seoul, Republic of Korea}
\acmPrice{15.00}
\acmDOI{10.1145/3387940.3392241}
\acmISBN{978-1-4503-7963-2/20/05}

\begin{document}

\title[More than Code: Contributions in Scrum Software Engineering Teams]{More than Code:\\Contributions in Scrum Software Engineering Teams}

\author{Frederike Ramin}
\email{frederike.ramin@student.hpi.de}
\orcid{}
\affiliation{%
  \institution{Hasso Plattner Institute}
  \city{University of Potsdam, Germany}
}

\author{Christoph Matthies}
\email{christoph.matthies@hpi.de}
\orcid{0000-0002-6612-5055}
\affiliation{%
  \institution{Hasso Plattner Institute}
  \city{University of Potsdam, Germany}
}

\author{Ralf Teusner}
\email{ralf.teusner@hpi.de}
\orcid{0000-0002-5606-8819}
\affiliation{%
  \institution{Hasso Plattner Institute}
  \city{University of Potsdam, Germany}
}

\renewcommand{\shortauthors}{Ramin, Matthies, Teusner}

\begin{abstract}
Motivated and competent team members are a vital part of Agile Software development and make or break any project's success.
Motivation is fostered by continuous progress and recognition of efforts.
These concepts are founding pillars of the Scrum methodology, which focuses on self-organizing teams.
The types of contributions Scrum development team members make to a project's progress are not only technical.
However, a comprehensive model comprising the varied contributions in modern software engineering teams is not yet established.
We propose a model that incorporates contributions of all Scrum roles, explicitly including those which are not directly related to project artifacts.
It improves the visibility of performed tasks, acts as a starting point for team retrospection, and serves as a foundation for discussion in the research community.
\end{abstract}



\maketitle

\section{Introduction}
Modern software engineering features collaborative, iterative development by teams following development processes and practices customized to their project contexts.
The Agile Manifesto emphasizes the importance of teams, stating that ``the best [\ldots] designs emerge from self-organizing teams''~\cite{fowler2001agile}.
Agile methods, based on these principles, such as Scrum, have become the de facto standard in professional software engineering~\cite{ScrumAlliance2018,StateOfAgile2019}.
These processes highlight the importance of human factors~\cite{Matthies2019ICSE_SEET}.
They focus on teams that are cross-functional and which include team members with all the capacities and competencies required to accomplish the project work~\cite{Schwaber2017}.
Therefore, the types of contributions that software engineers make to the progress and success of modern software projects are varied.
They not only include technical aspects, e.g., source code changes, but also process improvement activities, meeting facilitation, and effective communication with colleagues~\cite{Gousios.2008}.
For the remainder of this paper we rely on the following definition:

\boxbox{0.95}{\emph{Contribution}: Any activity, demanding human resources, that adds to the fulfillment of project goals, by adding value to the developed product or the (future) effectiveness of the team.}

Ford et al. characterized the actions of software engineers.
They list tasks such as learning, knowledge dissemination, feedback, and networking, which are vital to team success~\cite{Ford2017}.
Recent studies of software developers found that coding-related activities only took up one-fourth of their total work, with another fourth being used for collaborative activities~\cite{Meyer2019}.
The Scrum framework acknowledges these different task profiles, proposing specific roles, i.e. \emph{Product Owner} (PO), \emph{Scrum Master} (SM) and \emph{Development Team} (Dev.), with distinct responsibilities~\cite{Schwaber2017}.

Agile approaches rely on clear communication and visibility of progress to enable efficient collaboration and effective teams~\cite{Kniberg2015}.
Capturing and categorizing the contributions of team members is a cornerstone of ensuring team awareness regarding accomplished work.
It enables the appropriate valuation and appraisal of contributions to team efforts necessary for successful teamwork.

\begin{figure*}[ht!]
  \centering
  \includegraphics[width=0.7\linewidth]{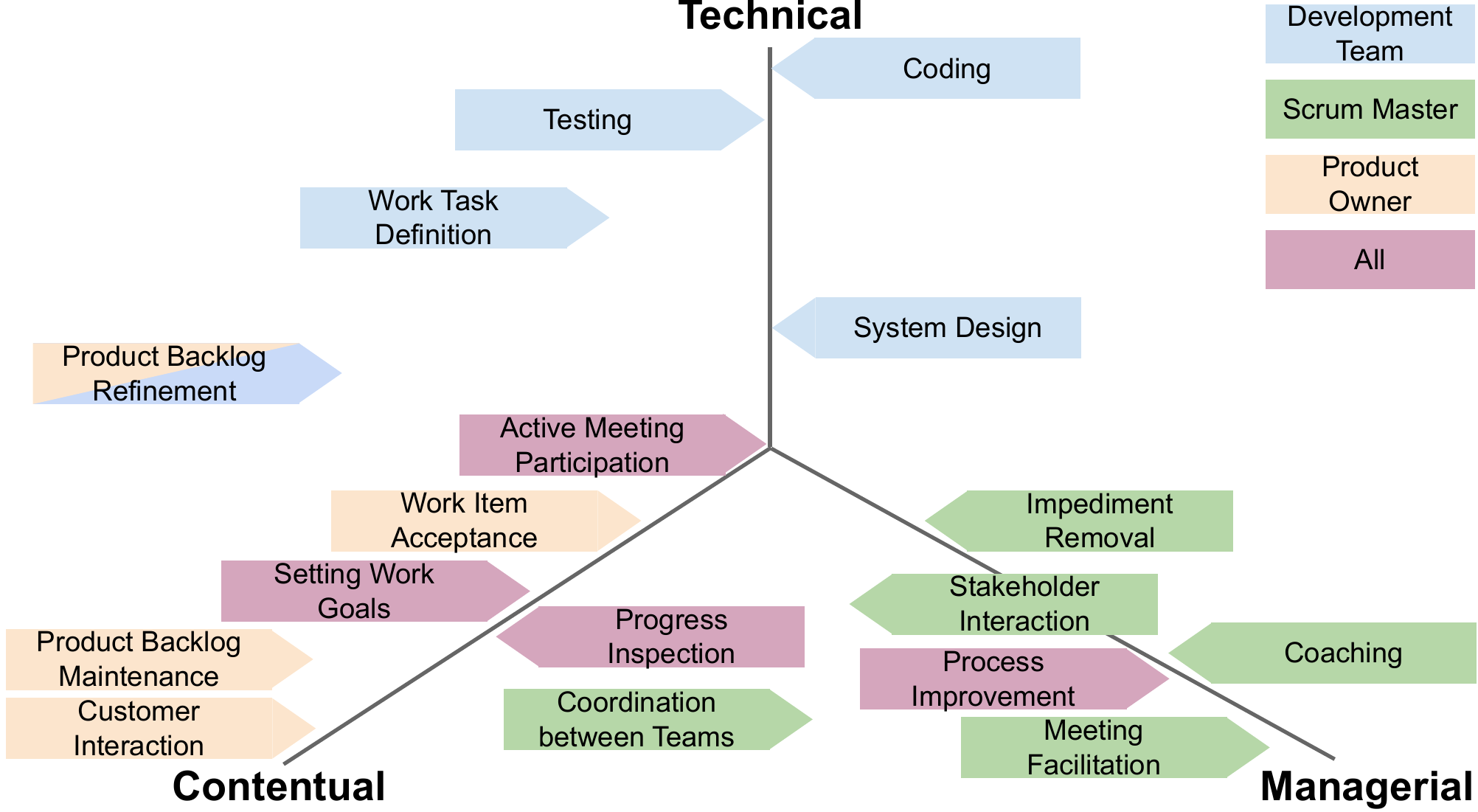}
  \caption{Visual representation of our initial contribution model regarding contributions to Scrum teamwork by participant roles (color coded). The closer a contributions is placed at the ends of a dimension, the more it is related to this category.}
  \Description{}
  \label{fig:scrum_model}
\end{figure*}

\section{Background}
The characterization of contributions to software projects, and their assessment, is an ongoing field of research.
Previous work in this domain includes research on models of software engineering \emph{tasks}, i.e. specific activities, and the personality traits of developers~\cite{Glass.1992,Wiesche.2014}.
These studies are predominantly concerned with traditional software engineering approaches, categorizing contributions by the different phases of software development or the roles that perform them~\cite{S.Sodiya.2007,Wiesche.2014}.
Additional previous work focused on the technical aspects of software engineering~\cite{Gousios.2008}, dividing engineer's tasks into activities such as IT support and database administration~\cite{Snyder.2006} or compilation and debugging~\cite{Singer2010}.
While multiple models of teamwork have been proposed, the activities, tasks, and responsibilities specific to modern, Agile software engineering teams have not been the core focus of recent studies.

In 1989, Goldstein classified software project team members according to the tasks they regularly perform~\cite{Goldstein.1989}.
He identified four distinct groups: \emph{programmers}, \emph{analysts}, \emph{maintainers}, and \emph{supporters}, each contributing in different ways.
Similarly, Glass et al. presented an early model of software engineering contributions, listing tasks such as debugging, reviews or user training~\cite{Glass.1992}.
The authors tag these activities as either routine or non-routine.
They conclude that in the domain of software engineering the intellectual, i.e. non-routine, tasks dominated routine tasks by a ratio of almost 4 to 1.
More recently, Gousios et al. have pointed out, that, even in the field of Software Repository Mining, ``no clear definition'' of software engineering contributions existed~\cite{Gousios.2008}.
However, the authors state that source code should not be the only contribution metric, especially in the context of Agile development.
They identified technical contributions through a hierarchical, top-down approach, based on project assets and the actions that can be performed on them.
Nonetheless, contributions not directly resulting in asset changes, e.g. facilitating meetings, are disregarded in this model.

\section{A Model for Contributions in Scrum}
Little previous research on contributions to teamwork has focused on the human aspects that Agile methods and the Scrum process framework stress.
We, therefore, constructed an initial model from first principles, based on the seminal Scrum Guide~\cite{Schwaber2017} by Ken Schwaber and Jeff Sutherland, the originators of the method.
We successively coded the guide's text, particularly focusing on the sections dealing with Scrum roles, artifacts, and meetings.
For every paragraph, we extracted the passages which mentioned work items, project requirements and responsibilities that the process stipulates for the different Scrum participants.
These items, after deduplication and clustering, represent the contributions of Scrum team members to the development process.
For each identified contribution, we assigned a short name and noted the topic as well as which role was designated for it.
\Cref{tab:extracts} contains examples of this process.

\begin{table}
  \caption{Examples of annotated Scrum Guide~\cite{Schwaber2017} extracts}
  \label{tab:extracts}
  \begin{tabularx}{\columnwidth}{Xll}
    \toprule
    \textbf{Text} & \textbf{Name} & \textbf{Role}\\
    \midrule
    ``The SM serves the Development Team [\ldots], including: [\ldots] Removing impediments to the Development Team'' &  \multicell{Impediment\\Removal} & SM \\
    \cmidrule(lr){1-3}
    ``The Development Team consists of professionals who do the work of delivering a [\ldots] product at the end of each Sprint.'' & Coding & Dev.\\
    \cmidrule(lr){1-3}
    ``Each [Product] Increment is [\ldots] thoroughly tested, ensuring that all Increments work together.'' & Testing & Dev.\\
    \bottomrule
  \end{tabularx}
\end{table}

\Cref{tab:contribs} presents the 17 individual Scrum teamwork contributions we extracted, assigned to the Scrum roles: \emph{Development Team}, \emph{Scrum Master} and \emph{Product Owner}.
We explicitly included the role of \emph{All}, to highlight contributions that involve a high level of collaboration between all roles involved in the development process.

\begin{table*}
  \ra{1.1}
  \caption{Overview of the 17 Scrum project work contributions included in the model, grouped by roles. Based on~\cite{Schwaber2017}.}
  \label{tab:contribs}
  \begin{tabularx}{\textwidth}{llX}
    \toprule
    \textbf{Name} & \textbf{Roles} & \textbf{Description}\\
    \midrule
    Coding & Dev. & Producing a product increment which satisfies the Sprint Goal\\
    Testing & Dev. & Writing and executing software tests, ensuring that product increments work together\\
    Work Task Definition & Dev. & Creating implementation tasks from Sprint Backlog items and a plan for delivering them\\
    System Design & Dev. & Planning the software architecture and fundamental structure of the product\\
    Product Backlog Refinement & Dev., PO & Adding details and estimates to work items and adjusting their priorities \\
    \cmidrule(lr){1-3}
    Product Backlog Maintenance & PO & Modifying the Product Backlog by adding work items and their priorities\\
    Customer Interaction & PO & Extracting product requirements and prioritization by communicating with the customer\\
    Work Item Acceptance & PO & Checking the produced product increment for compatibility with the product vision\\
    \cmidrule(lr){1-3}
    Meeting Facilitation & SM & Scheduling and leading through meetings, preparing an agenda and keeping the time-box\\
    \multicell{Impediment Removal} & SM & Resolving identified problems that hinder project progress, improving team workflows\\
    Stakeholder Interaction & SM & Collecting feedback from stakeholders and changing the interactions with the outside world\\
    Coaching & SM & Leading and passing on process knowledge to team members\\
    Coordination between Teams & SM & Communicate and distribute work between teams in multi-team settings\\
    \cmidrule(lr){1-3}
    Setting Work Goals & All &  Structuring the work of the next iteration (Sprint Planning) or the next day (Daily Scrum) \\
    Active Meeting Participation & All & Being involved and contributing to the meeting goals, supporting a positive attitude\\
    Process Improvement & All & Inspecting and adapting the employed process (Sprint Retrospective)\\
    Progress Inspection & All & Examine the work performed in the latest product increment\\
    \bottomrule
  \end{tabularx}
\end{table*}

We relied on previous work in the area of project management to identify the main traits and topic areas that teamwork contributions are classified by.
Ebert and de Man~\cite{Ebert.2008} grouped software engineering knowledge into the three areas: \emph{Project} (e.g. requirements, budget, timing, milestones), \emph{Product} (e.g. product features, relations to other products or standards) and \emph{Process} (e.g. business processes, workflows, responsibilities).
Similarly, Wynekoop and Walz~\cite{Wynekoop.2000} divided the traits of top-performing software developers into three related categories: (i) traits of those who best ``make things work'', (ii) traits of those who best communicate with users, and (iii) traits of those ``destined for management''.
We employ a combination of these categories as dimensions to characterize contributions, based on the language of the Scrum Guide:
\begin{itemize}
    \item \textbf{Technical}: How much does the contribution add to the product increment, utilizing technical knowledge and skills?
    \item \textbf{Contentual}: To what degree does the contribution influence the product's prospect and direction?
    \item \textbf{Managerial}: To what degree is the contribution concerned with adapting the work process, not the product?
\end{itemize}
These three model dimensions are non-exclusive.
Project contributions may involve technical, contentual and managerial aspects simultaneously, though to different extents.
For example, a Scrum Master solving identified issues as part of \emph{Impediment Removal} is most likely dealing with adapting work processes.
However, depending on the specific issue, these changes may also require some technical skills or may affect the developed product and how it is built\footnote{Conway's Law: ``organizations which design systems [\ldots] are constrained to produce designs which are copies of the communication structures of these organizations.''~\cite{Conway1968}}.
The proposed model dimensions thus create a gravity field to locate and categorize project and teamwork contributions.

Figure~\ref{fig:scrum_model} presents our visualized model of contributions to project work in Scrum teams.
The closer a specific contribution is to the edges of the graph, the more it is related to only one or two dimensions.
For example, the contribution of \emph{Coding}, i.e. producing source code and ``delivering a potentially releasable increment''~\cite{Schwaber2017}, is at the far end of the \emph{Technical} dimension.

\section{Discussion}
The proposed model represents a structured exploration of the project member's contributions described by the Scrum Guide~\cite{Schwaber2017}.
We explicitly constricted the scope of this initial model to promote clarity and traceability of construction.
The model represents the state of the Scrum Guide in its latest version of November 2017.
The related literature on Scrum is vast and contains many more collections of tasks and roles of team members in the software engineering domain~\cite{Kniberg2015,Ford2017}.
Since the Scrum process model needs to be adapted to the context in which it is used, as well as to the team which uses it, many different implementations exist in practice.
As such, the presented model serves as a basic structure, representative of key contributions in unmodified, theoretical, ``vanilla Scrum''~\cite{Gren2017}.
The Scrum Guide itself lists adaptation as one of the core elements of the underlying theory and states that ``specific tactics for using the Scrum framework vary and are described elsewhere''~\cite{Schwaber2017}.
While Scrum contains prescriptions for team organization, it contains few specifics on how software development activities should be performed.
Our model, therefore, reflects this approach.
In practice, Scrum is often employed in conjunction with additional, complementary methods, most notably XP, which suggests specific tactics on how work is to be performed, e.g. Pair Programming~\cite{Kniberg2015}.

Future work will focus on this aspect, including contributions from additional sources,  further Agile methods, and industry practice in the model.
The base model presented here can then also be employed to show differences between the project contributions expected in different Agile methods.
Furthermore, we envision the proposed model to be helpful in software engineering projects, e.g. in the following use cases:

\paragraph{Scrum Team Status Check}
The model can be used as a means of conformance analysis, contrasting a team's process and self-identified teamwork contributions to those designated by the Scrum Guide.
In an initial step, after identifying their own roles, team members can investigate whether they make all of the contributions listed for their specific role.
Any mismatches in this process represent starting points for discussion in the team.
This allows retrospection on the chosen process adaptations and their rationales.

\paragraph{Team Contribution Analysis}
The list of Scrum team contributions, see \cref{tab:contribs}, can be reviewed,
discussing for every item whether it is clear who currently is or should make a specific contribution.
This can identify project contributions that have previously been overlooked in a team.

\paragraph{Contribution Awareness}
The proposed teamwork model contains many contributions of an interpersonal and non-technical nature, which are core to Scrum but receive less focus in more traditional software development approaches.
A visual representation of these contributions as presented in \cref{fig:scrum_model}, can improve developers' awareness of their own behavior~\cite{Meyer2018} and their contributions besides writing code.

\paragraph{Agile Process Coaching}
The Scrum contributions to project work represent the daily activities and responsibilities of Scrum team members.
They define what a developer, Scrum Master, or Product Owner will likely spend a significant portion of their work time on.
While the proposed model focuses solely on the Scrum process framework, it can be used to highlight the differences in work activity between different Agile methods.
By highlighting which contributions would be impacted by changes in the development process flow, shifts in daily work can be anticipated.
This approach thus allows an easier transition from one method to another by focusing on the changes in daily contributions that need to happen.

\medskip
Our model makes the often implicit contributions of Scrum team members to project progress explicit.
It allows comparisons of one's own team state to the ``by the book'' Scrum process.
These types of analyses are suitable to foster self-reflection and retrospection regarding teamwork processes and contributions in teams.
They may, therefore, prove particularly useful in \emph{Retrospectives}, Scrum's implementation of software process improvement~\cite{Matthies2019ICSE_DS}.
In Retrospective Meetings, the team reflects on what aspects of the last development iteration were beneficial and should be kept, and what aspects should be improved in the future~\cite{Schwaber2017}.
To structure these meetings, encourage active participation and to break the usual routine, interactive \emph{Retrospective Games} have been introduced by Agile practitioners~\cite{Derby2006}.
These games often make use of brainstorming, visualizations or metaphors to generate process improvement ideas.
The proposed model of Scrum project contributions can provide a default view of Scrum for teams to compare themselves against.
It complements the array of existing tools to facilitate productive team reflection and retrospection.

\section{Conclusion}
Software development project teams are often characterized as homogeneous groups, without taking the varied tasks and roles of team members into account~\cite{Ford2017}.
However, contributions to a software development project are not constricted to coding, with developers routinely spending only about half of their workday working on their computers~\cite{Meyer2019}.
This fact is especially relevant for teams employing the Scrum process framework, which highly depends on effective self-organization, communication, and collaboration~\cite{Matthies2016c}.
This reality of modern software engineering teams is reflected only inadequately in previous models of teamwork contributions.
We, therefore, present a model based on project contributions defined in the Scrum Guide~\cite{Schwaber2017} to address this research gap.
This model, categorizing the teamwork contributions of Scrum roles using the dimensions \emph{Technical}, \emph{Contentual} and \emph{Managerial}, is an initial proposal.
It is open to refinement, discussion, and enrichment using additional sources and Agile method descriptions.
Our model fosters retrospection and represents a first step towards a more complete view and understanding of the types of contributions Scrum team members make to successful projects.


\bibliographystyle{ACM-Reference-Format}
\bibliography{main}

\end{document}